# Near-field terahertz probes with room-temperature nanodetectors for subwavelength resolution imaging

Oleg Mitrofanov,[1] Leonardo Viti,[2] Enrico Dardanis,[2] Maria Caterina Giordano,[2] Daniele Ercolani,[2] Antonio Politano,[3] Lucia Sorba,[2] and Miriam S. Vitiello [2*]

[1] *University College London, Electronic and Electrical Engineering, London, WC1E 7JE, UK*
[2] *NEST, Istituto Nanoscienze – CNR and Scuola Normale Superiore, Piazza San Silvestro 12, Pisa, I-561273*
[3] *Università degli Studi della Calabria, Dipartimento di Fisica, via Ponte Bucci, 87036 Rende (CS), Italy*

**Near-field imaging with terahertz (THz) waves is emerging as a powerful technique for fundamental research in photonics and across physical and life sciences. Spatial resolution beyond the diffraction limit can be achieved by collecting THz waves from an object through a small aperture placed in the near-field. However, light transmission through a sub-wavelength size aperture is fundamentally limited by the wave nature of light. Here, we conceive a novel architecture that exploits inherently strong evanescent THz field arising within the aperture to mitigate the problem of vanishing transmission. The sub-wavelength aperture is originally coupled to asymmetric electrodes, which activate the thermo-electric THz detection mechanism in a transistor channel made of flakes of black-phosphorus or InAs nanowires. The proposed novel THz near-field probes enable room-temperature sub-wavelength resolution coherent imaging with a 3.4 THz quantum cascade laser, paving the way to compact and versatile THz imaging systems and promising to bridge the gap in spatial resolution from the nanoscale to the diffraction limit.**

**Introduction.**

Imaging in the terahertz (THz) spectral range (frequency: 0.3-10 THz, wavelength: 30-1000 µm) is severely restricted by diffraction.[1,2] Thus near-field scanning probe microscopy methods are commonly employed to enable mapping of the THz electromagnetic fields with sub-wavelength spatial resolution,[3-11,12-20] overcoming the diffraction limit by up to 4 orders of magnitude.[7] Scattering-type scanning THz near-field microscopy[7,12] has been leading in spatial resolution by exploiting an atomic force microscope (AFM) tip to induce strongly concentrated THz fields at the tip apex. A fraction of the concentrated field is scattered, and the scattered wave carries information about the dielectric constant $\varepsilon$ underneath the tip in its amplitude and phase, therefore enabling spatial resolution determined by the tip apex size rather than the wavelength. However, the scattering efficiency of the AFM tip is prohibitively low in the THz frequency range, thus requiring the use of sophisticated signal demodulation techniques,[7,12] powerful THz lasers, cooled bolometers and costly THz time-domain spectroscopy systems to detect the weak waves scattered only from the tip apex. Furthermore, the atomic force interaction required to control the tip position restricts application of the scattering-type microscopy for life sciences. Conversely, the original near-field microscopy concept,[10] which overcomes the diffraction limit by detecting light through a sub-wavelength aperture in a metallic screen, relies neither on the scattering process nor on the atomic force interaction. The spatial resolution of this method is determined by the aperture size ($a$). In practice, however, the resolution is limited by the strong reduction of light transmission ($T$) through the aperture,[2] which was found by Bethe and Bouwkamp to follow a power law: $T \sim a^6$.[1,21] The sixth power dependence makes the use of apertures smaller than ~1/100 of the wavelength impractical for standard detection arrangements. Nevertheless, recently developed THz nano-detectors [22-32] offer a promising alternative solution.

The Bethe and Bouwkamp dependence applies to the propagating waves, whereas the transmitted field also contains evanescent waves, which are localized in the near-field zone of the aperture.[18] These evanescent waves represent the major component of the transmitted field for deeply sub-wavelength apertures (<$\lambda$/100).[9,11] However, being localized at the aperture, the evanescent fields remain undetected in most near-field imaging systems. Here, we introduce a novel near-field THz probe concept, where the evanescent THz field is converted into a detectable electrical signal at the nanoscale. We integrate a THz nanodetector based on a semiconductor nanowire (NW) or a thin flake of crystalline black phosphorus (BP) into the evanescent field region of a sub-wavelength aperture to enable efficient detection of the transmitted wave. This THz near-field probe provides sub-wavelength resolution at room temperature (RT) in a compact THz imaging system without the need for the mode-locked table-top femtosecond pulse lasers, atomic force interaction or demodulation techniques. We exploited a compact 3.4 THz quantum cascade laser (QCL) operating in pulsed mode with an average output power of 45 µW for a proof-of-principle demonstration of a practical and compact system for coherent THz near-field imaging. The devised near-field probe concept can exploit the growing variety of nanoscale THz detectors, including inexpensive silicon-based THz detectors,[33] and open a new route for enabling THz imaging and spectroscopy beyond the diffraction limit.

**Device concept**

Recently, single semiconducting NWs,[22,24,25] graphene flakes,[26] quantum dots,[27] and heterostuctures based on flakes of crystalline BP [23,28-29] have been successfully demonstrated for detection of THz-frequency waves. In particular, field-effect transistor (FET) architectures have been utilized for detection in the 0.3-3 THz range with less than 100 pW/Hz$^{1/2}$ noise equivalent power (NEP)[24] and signal-to-noise ratio as high as 20000.[29] The 1D InAs NWs provide a clear benefit for the development of detectors with fast read-out, due to the inherently small atto-farad level capacitance;[23] whereas the 2D geometry of the atomically thin BP and graphene FETs is particularly promising for the detection of highly localized evanescent fields.

Asymmetric feeding of THz radiation into the FET channel, which produces the THz-frequency field between the gate (G) and the source (S) electrodes can enable RT THz detection via a number of different physical phenomena.[28] In the resistive self-mixing rectification effect,[30] the THz field modulates the charge density ($\Delta n$) in the channel in phase with the longitudinal driving field. At RT an overdamped collective density oscillation of carriers is formed and a steady-state source-drain current $i_{SD}$ or voltage $\Delta u_{SD}$ develops across the FET channel.[30] FETs can also detect THz radiation very efficiently by means of the thermoelectric effect [28] as a consequence of the temperature gradient ($\Delta T$) along the channel, caused by the asymmetric feeding of the incident THz wave to the FET electrodes. Under this condition, a steady-state thermoelectric voltage $\Delta u_{SD} = \Delta T \Delta S_b$ (where $S_b$ is the Seebeck coefficient) develops across the channel. The voltage is proportional to the incident THz wave intensity. [28]

To feed the THz field, incident on a sub-wavelength aperture, to the FET channel asymmetrically, we engineered a special gate (G) electrode, which simultaneously acts as a metallic screen with a sub-wavelength aperture. The electrode extends asymmetrically from one side of the aperture to the FET placed in the aperture center (Fig. 1a). An electrically isolated S electrode in the same trapezoid shape extends to the aperture center from the opposite side. When exposed to the incident THz field, these electrodes concentrate the field in the FET channel (see Fig. 1b). The drain (D) electrode is fully covered by the G electrode (Fig. 1a) and kept electrically isolated from it by means of a dielectric ($SiO_2$) layer. This aperture geometry breaks the symmetry of the S and D electrodes and results in the alternating THz field between G and S that activates the THz detection mechanisms.

A set of near-field probes was fabricated using single NWs with diameters in the range of 50-100 nm or 10 nm thick BP flakes. We lithographically defined a 1.1 μm long SD channel for both the NW sample (Sample A, Fig. 1c) and the BP sample (Sample B, Fig.1d). For sample B, the BP flake was contacted along the armchair (*x*) crystalline direction, *i.e.* the direction of higher carrier mobility, and then encapsulated by a 80 nm thick layer of $SiO_2$ to avoid degradation under oxygen exposure.[28] This layer was also utilized as the FET gate dielectric. A thicker (120 nm) SiO$_2$ layer was required for the NWs in order to prevent current leakage between the channel and the G electrode. An aperture was formed around the THz detectors by depositing a 10/150 nm thick Cr/Au screen on top of the dielectric layer. The field-concentrating trapezoid extension, acting as the G electrode, was aligned with the S electrode at a S-G distance of 500 nm (see Methods).

**Results and discussion**

The response of the aperture-coupled THz nano-detectors was tested using a pulsed QCL operating at 3.4 THz (see Methods). Figure 2a shows a typical source-drain current $i_{SD}$ at the output of the BP-based near-field probe with a 20x20 µm$^2$ aperture as a function of the incident THz beam intensity. The current increases linearly with the beam intensity (Fig. 2a). Alternatively, the incident THz wave can induce a SD voltage $\Delta u_{SD}$ if the detector is operated in an open circuit configuration; $\Delta u_{SD}$ increases linearly with the QCL power as well (Fig. 2a). Similar linear output characteristics were also found for sample A (Fig. 2a).

In order to evaluate the photoresponse of sample A we estimate the power incident on the entire aperture (including the area of the electrodes). First, sample A (with a 20x20 µm$^2$ input aperture) was scanned perpendicular to the THz QCL beam to map the beam profile in a focal plane. The intensity profile closely follows a Gaussian distribution with the beam radius of $r$ = 83.5 µm, which corresponds to the beam full width at half maximum of FWHM = 2.355$r$ = 197 µm (Fig. 2b). Knowing the total THz power and the profile of the THz beam, we can determine the power density at the beam center: $I_{THz} = P/(\pi r^2)$, where $P$ is the total average power in the THz beam. As the aperture is several times smaller than the beam diameter, we estimate that the detector receives a total amount of incident power $P_A = I_{THz} S_A$, where $S_A$ is the aperture area ($S_A$ = 20x20 µm$^2$), if the aperture is placed in the beam center. Based on these considerations, we found that the responsivity of sample A at $V_G$ = 0 V is $R_{nw}$ = 2.5 V/W (see Methods), more than a factor of two larger than that achieved in THz NW detectors operating at lower frequencies (2.8 THz) in the standard far-field architecture.[31] For sample B, the unbiased responsivity is one order of magnitude lower ($R_{BP}$ = 0.17 V/W).

The near-field probe produces a signal proportional to the THz field intensity averaged over the aperture area and therefore enables mapping of the THz intensity distribution with a spatial resolution determined by the aperture size. The intensity mapping, however, provides only partial information. Most THz near-field imaging applications also require phase information.[4,32] For this purpose, it is convenient to set the incident wave with the same phase across the image area. The latter condition can be realized by orienting the probe surface perpendicularly to the incident THz QCL beam. In this case, the metallic screen of the probe and the QCL output facet form a Fabry-Perot cavity of length $L$ (Fig. 3a). The metallic surface reflects the incident THz wave back to the QCL, forming a standing wave if $L$ is equal to an integer multiple of the half-wavelength $\lambda_{THz}/2$. We verified the standing wave formation by translating the probe along the axis and therefore adjusting $L$. Figure 3b shows a map of the THz field detected while varying the position of the sample B probe in the $yz$-plane. The intensity distribution along the $y$-direction follows the profile of the incident THz beam, similar to the profile in Fig. 2(b), whereas the field along the $z$-direction shows regular fringes occurring with a period of ~ 44±2 μm, which corresponds to a standing wave for the QCL wavelength of $\lambda_{THz}$ = 88 μm. By translating the probe by $\lambda_{THz}/2$ along the $z$-axis from one fringe to the next, the THz field in the cavity reduces below the detection level (between the fringes). The fringes are perpendicular to the $z$-axis, indicating that the incident THz field in the image plane indeed has the same phase.

Although the interference fringes can be eliminated completely by tilting the surface normal away from the $z$-axis (Fig. 3c), the formation of the standing wave with regular fringes in the *yz*-map suggests that the normal incidence configuration can be exploited for phase-contrast THz imaging and spectroscopy. As a practical example one can consider a sub-wavelength size THz resonator, such as a micrometer-scale dielectric sphere,[3] illuminated by a THz wave of frequency close to the frequency of the Mie magnetic dipole mode. If the resonator is positioned in front of the aperture, the THz field coupled into the aperture is a superposition of the standing wave field and the polarization induced by the standing wave in the resonator. It has been shown that the detected near-field signal is strongly enhanced in the case of constructive interference of the polarization field with the standing wave or fully suppressed in the case of destructive interference.[3] Therefore, the phase of the induced polarization, which varies rapidly with frequency near the resonance, affects the signal detected by the near-field probe. This effect promises phase-sensitive THz imaging, which can be particularly powerful if a widely tunable QCL is employed [34,35] to vary the frequency and thus the phase of the induced polarization.

In a more general approach, the phase of the detected field in the present probe architecture can be measured interferometrically by combing the field coupled through the aperture with a reference THz wave from the same QCL, incident on the probe from the back side though the Si substrate. In such a configuration, the FET detector produces a photovoltage signal $\Delta u_{SD}$ proportional to intensity of the two superimposed waves in the region of the FET channel: $\Delta u_{SD} \sim (E_1 + E_{REF})^2$, where $E_1$ and $E_{REF}$ are the local electric field amplitudes due to the wave coupled through the aperture and the reference wave, respectively.

Interferometric phase measurements were performed at the center of a focused THz beam from the QCL, by using sample A. First, we evaluated photoresponse of the probe to the THz wave incident through the substrate: $\Delta u_{SD}$ was approximately four times higher compared to the photovoltage produced by the same THz wave passing through the aperture.

The power density in the reference beam therefore was chosen to be approximately four times lower in comparison to the beam incident on the aperture in order to produce similar photovoltages for each of the beams. When these two counter-propagating beams impinge on the probe, we expect that constructive interference of the two beams produces coherent gain and a photovoltage four times higher the photovoltage arising from individual beams. We note that in imaging applications, the reference wave is likely to have a negligible effect on samples because it is blocked from reaching the sample area by the metallic screen of the probe.

In the experiment, sample A was illuminated from two sides using two beams from the same QCL. Details of the experimental set-up can be found in Supplementary Information. The beam was split using an undoped Si wafer (BS, Fig. 4, inset). The transmitted beam was focused on the aperture side using a parabolic mirror (P2); the reflected beam was focused using another parabolic mirror (P1). Relative phase of the two beam was adjusted using a delay stage with two mirrors (M1 and M3) aligned to form a 90 degree angle between their surfaces.

We observed enhanced photovoltage when the reference beam $I_{REF}$ is incident on the back of the near-field probe with a clear signature of interference when the relative phase of the reference beam was adjusted using the delay stage. The photovoltage varied periodically as the reference beam path was increased. The average period over a scan of 500 μm was $p = 43.5 \pm 0.5$ μm. It corresponds to the wavelength of $\lambda = 2p = 87\pm1$ μm (Fig. 4), coincident with the QCL wavelength. Maxima in the detected signal correspond to constructive interference of the field coupled through the aperture and the field of the reference beam.

To ensure that no standing wave is formed between the probe and the QCL, the delay stage position was varied while only the reference beam illuminated the probe. No periodic intensity variation was observed (Fig. 4, blue line). Furthermore, no indication of a standing wave was observed when the probe was translated along the optical axis while one of the beams was blocked. It is worth mentioning that the distance between the QCL and the probe in the phase experiments was ~ 40 cm. A larger distance compared to the experiment presented in Figs. 3 makes formation of a standing wave more sensitive to the probe surface orientation.

To verify that the near-field probe described here also provides a spatial resolution comparable with the aperture size (~20 μm), we mapped THz intensity distribution formed at the tips of two metallic needles separated by ~10 μm and placed in the focus of the THz beam (see Supplementary Informations) The intensity map showed clear confinement of the THz field between the needles with a full width at half maximum of ~17 μm.

To improve the spatial resolution of this near-field probe, the aperture size can be reduced without any changes in the nanodetector architecture. The output signal, in this case, however will decrease due to the strong dependence of the aperture transmission coefficient on the aperture size.[1,14] To evaluate the practical limit of the input aperture size, and consequently the spatial resolution achievable with the proposed architecture, we modeled the near-field probe using a commercial full-wave electromagnetic simulation software (COMSOL). Figure 5a shows the electric field intensity along the axis passing through the aperture center for aperture sizes of 20, 10, 5 and 2 μm. On the left hand side of the aperture plane (located at $z=0$), the intensity distribution varies periodically with $z$, corresponding to the interference between the incident and the reflected wave. On the right hand side (behind the aperture), the intensity drops rapidly. The decay is substantially faster than the $1/z^2$-dependence expected for the propagating waves, indicating the evanescent nature of the transmitted field in this region. Nevertheless, in the aperture plane ($z = 0$) and within a short distances from it, $z = -100$ nm, where the nanodetector is placed, the field intensity is enhanced by over one order of magnitude due to the field concentrating between the G and S electrodes. Figure 5b shows the normalized field intensity ($|E|^2$) at the location of the nanoscale detector ($z = -100$ nm) for apertures ranging from 1 to 20 μm in size. Although the plot shows a rapid drop in intensity following approximately an $a^4$ dependence, the decay at $z = -100$ nm is less severe than that at $z = 2$ μm, where the simulation results agree with the $a^6$ dependence predicted by Bethe and Bouwkamp (Fig. 5b).

Figure 5b provides a useful insight into the effects of the near-field probe architecture on the sensitivity. The field enhancement that yields the high sensitivity of the probe can be exploited only with detectors placed at ultra-short distances ($z$) from the aperture plane (see Fig. 5b), using a thin gate dielectric layer, while keeping the current leakage through the gate oxide negligible. The gap between the G and S electrodes that controls the THz field concentration in the FET channel also directly affects the probe sensitivity. Therefore, the atomically thin detectors, such as the BP flakes, are more favorable for this near-field probe architecture as they can be fabricated with a thinner gate oxide and a shorter channel compared to the 50 nm diameter NWs. The simulations in Fig. 5b also allow us to evaluate the potential of the probe with employed NW nanodetectors for higher spatial resolution THz imaging applications. In fact, the estimated SNR of 65 for the 20 μm aperture NW probe sets a limit for the smallest usable aperture: a probe equipped with an aperture of 5 μm would have a SNR of 2.

The simulations additionally show an enhancement in the THz field in the FET channel for apertures sizes of 10-20 μm. In fact, we observed similar output voltages $\Delta u_{SD}$ for near-field probes fabricated with 10 μm and 20 μm apertures. The aperture, in this case, exhibits the antenna effect and resonantly enhances the THz field in the area of the FET detector, and thus enhances the responsivity (Fig. 5a). The 10-μm aperture probes therefore can be used to enhance the dynamic range for THz imaging applications with moderate spatial resolution. Alternatively, THz nanodetectors coupled with 10-20 μm apertures can be easily implemented in an array architecture forming an imaging sensor with sub-wavelength size pixels designed for enhanced sensitivity for selected THz frequencies.

It is worth discussing the origin of the physical detection mechanisms activated in the aperture-embedded BP and NW detectors. Relevant mechanisms for samples A and B were identified using the observed gate voltage dependence of the photoresponse (Fig. 6a and 6c) and the related sign.[28] Sample A (Fig. 6a) exhibits minimum photoresponse $\Delta u_{SD} = 0$ V at $V_G = -0.25$ V and the maximum photoresponse at $V_G \sim 0$V. The photovoltage sign remains negative along the whole spanned gate voltage range, a clear signature of the photo-thermoelectric effect associated with the diffusive flux of electrons from the hot-side (S) to the cold-side (D) of the channel caused by the THz-induced carrier redistribution. The shape of the photovoltage curve qualitatively matches the predicted thermo-electric photovoltage,[28] which is proportional to the Seebeck coefficient $S_b$ [28] obtained from the Mott equation (see Supplementary Information). The maximum device responsivity, reached at $V_G = 0$ V, corresponds to a minimum noise equivalent power NEP $\sim 4.8$ nW/Hz$^{½}$ and to SNR $\approx 65$ (Fig. 6b)

The analysis of detection mechanisms in sample B (see Supplementary Information) allows us to conclude that the same thermoelectric dynamics dominate in the BP detector. Figure 6c shows the comparison between the photoresponse and the extrapolated Seeback coefficient (see Supplementary Information). Furthermore, the photovoltage sign remains positive along the whole spanned gate voltage range, a clear signature of the photo-thermoelectric effect associated with the diffusive flux of holes from the hot-side (S) to the cold-side (D) of the channel. Such a behavior is expected for the *x*-oriented BP flake due to the lowest acoustic photon speed along the armchair crystal axis. [28,29] The detector responsivity largely increases by tuning the gate voltage; in particular, a maximum responsivity of 1.3 V/W has been found for $V_G = 1.5$ V, corresponding to a minimum NEP of ~ 55 nW/Hz$^{1/2}$ (Fig. 6d).

**Conclusions**

In conclusion, we demonstrate a novel near-field probe architecture exploiting THz nanodetectors embedded in the aperture region as a successful and practical sensor for THz sub-wavelength resolution imaging. The incorporation of the nanodetector within the sub-wavelength aperture allows us to mitigate the severe attenuation of the transmitted wave by innovatively utilizing the evanescent components of the transmitted wave. This paves the way to high-resolution room-temperature THz imaging, which so far has been relying predominantly on THz detection techniques that require either an ultrafast laser or a cryogenically-cooled THz detector. In particular, the near-field probes discussed here can enable coherent near-field spectroscopy and imaging of multipolar plasmonic modes in spoof plasmon THz resonators, which typically have sharp resonances and complex mode structure [36]. Application of the aperture-type near-field probes was recently demonstrated for imaging and spectroscopy of spoof plasmon resonators using broadband THz time-domain spectroscopy technique [37]. The near-field probes discussed here can provide narrow-bandwidth characterization by employing THz QCLs. We demonstrate a proof-of-principle room-temperature application of the probes with embedded THz nanodetectors for coherent imaging using a 3.4 THz QCL. The growing field of nanoscale THz detectors, particularly the silicon based FET detectors, can improve the sensitivity and spatial resolution of this technique and deliver inexpensive room-temperature sub-wavelength resolution THz imaging technology.

**Acknowledgements**

The authors acknowledge support from the European Union ERC Grant SPRINT, the European Union through the MPNS COST Action"MP1204 TERA-MIR Radiation:

Materials, Generation, Detection and Applications" and the Royal Society [Grant No. UF 130493]

Methods

**Nanowire growth.** 2 μm long InAs NWs were grown bottom-up by chemical beam epitaxy via the Au-assisted growth using trimethylindium (TMIn) and tertiarybutylarsine (TBAs) as metal-organic (MO) precursors. Se-doping was used to control the charge density in the NW and to optimize the channel and contact resistances, while ensuring sharp pinch-off in the FET transconductance.[25] The NWs were then mechanically transferred onto a 350 μm thick high-resistivity Si substrate with a 300 nm insulating $SiO_2$ layer.

**Black phosphorus.** Single-crystalline ingots of BP were grown via a chemical vapor transport technique.[28] Red phosphorus was placed into an evacuated quartz tube in a double-zone tube furnace with temperature set at 600 °C and 500 °C for the hot and cold end, respectively. Large-size single crystals of BP were obtained after one week of vapor transport. Flakes were mechanically exfoliated from the bulk BP crystal using the adhesive tape technique and transferred onto the $Si/SiO_2$ substrates.

**Nanofabrication.** AFM mapping has been performed to evaluate the BP flake thickness by employing a Bruker system (IconAFM) with thickness resolution < 1 nm and lateral resolution ~20 nm. Each flake shows a layer thickness integer multiple of ~ 0.61 nm, i.e. the thickness of a BP monolayer (phosphorene). The in-plane orientation of BP flakes has been determined via micro-Raman spectroscopy. The BP flakes were individually contacted by 10/70 nm Ni/Au S and D electrodes via electron beam lithography (EBL) and thermal evaporation. The orientation of the FET channel for the BP flakes was chosen along the armchair direction. The InAs NW was contacted by 10/100 nm Ti/Au S-D electrodes. The $SiO_2$ encapsulating layer was then deposited on BP sample via Ar sputtering. This is a crucial step to prevent degradation due to ambient air that alters the electrical device performances and deteriorate the flake itself. The NW samples were covered with 120 nm of $SiO_2$ oxide layer due to the larger lateral size of the InAs NWs (50 nm diameter). The optimal oxide thickness was assessed by fabricating a set of test samples for which the leakage channel through the gate electrode was determined. The G electrode was aligned with the center of the channel via EBL and defined via thermal evaporation of 150 nm layer of Cr/Au. The BP and NW FETs showed Ohmic behavior with typical SD resistivities of 20 and 10 kΩ at $V_G$ = 0 V for BP and NW devices, respectively.

**Optical Characterization.** The QCL beam was focused on the aperture area positioned at $L$ = 15 cm using a pair of Picarin lenses having focal lengths of 2 cm (lens facing the QCL) and 5 cm (lens facing the detector), respectively. The QCL was driven by 500 ns current pulses at 40 kHz at the operating temperature of 26-28 K. The output intensity of the QCL was varied between 0-50 μW by adjusting the bias from -45 to -48V and monitored by a pyroelectric detector. The QCL current was modulated by a square function at 1.333 kHz and 50% duty cycle and the THz detector response was measured with a lock-in amplifier (detection bandwidth of 1 Hz). The detector responsivity was measured in the photocurrent and photovoltage configurations. In the latter, the S electrode was grounded, $V_G$ was set with a Keithley *dc* generator and $\Delta u_{SD}$ was measured at the D electrode with a low-noise voltage pre-amplifier (DL Instruments 1201, input impedance = 10 MΩ, pass-band filter: 300 Hz - 3 kHz, gain $G_n$ = 100) and a lock-in amplifier (SRS 5210). $\Delta u_{SD}$ was calculated using the lock-in amplitude *LIA*:

$$\Delta u = \frac{\sqrt{2}\frac{\pi}{2} \cdot LIA}{G_n}$$

The factor π√2/2 is a normalization coefficient that takes into account the *RMS* value of the fundamental sine wave Fourier component of the square wave produced by the current source. The responsivity ($R_v$) was then determined using the relation $R_v = (\Delta u_{SD}/P_A)$. In the photocurrent configuration, a low-noise current amplified (DL Instruments 1211) is used instead of the voltage pre-amplifier. From the responsivity value, it is possible to evaluate another figure of merit of the detector: the noise equivalent power (NEP), defined as the ratio between the noise spectral density (N) and the responsivity ($R_v$). Here, N is typically approximated as the Johnson Nyquist noise, induced in the FET channel by thermal fluctuations: $N = (4 k_B R_{ch} T_{RT})^{1/2}$, where $k_B$ is the Boltzmann constant, $T_{RT}$ is the temperature and $R_{ch}$ is the total channel resistance.

**Numerical Simulation.**

The coupling of the incident THz wave through the aperture of the near-field probe was simulated using a commercial Software (COMSOL Multiphysics). In the model, the aperture size was varied from 1 μm to 20 μm. A plane wave of power 1 W is sent on the detector at normal incidence. The resulting spatial distribution of the electromagnetic field was computed by using a finite elements approach (FEM) and the power density was calculated point-by-point in the entire simulation volume.

**Author Contributions**

O.M. and M.S.V. conceived the architecture and the experiments. O.M, M.S.V. and L.V. devised the near-field probes. L.V. and E.D. fabricated the samples. D.E., L.S. and A.P. performed the material growth and characterization. L.V., E.D. M.C.G and O.M. performed the transport and optical experiments. L.V., O.M, E.D. M.C.G. and M.S.V. analyzed and modeled the data. O.M. and M.S.V. co-wrote the paper. M.S.V. supervised the project.

**Competing financial interests:** The authors declare no competing financial interests.

**Materials & Correspondence**

Correspondence and requests for materials should be addressed to M.S.V. (miriam.vitiello@sns.it) and L.V. (leonardo.viti@nano.cnr.it)

**References**

1. Bethe, H. Theory of diffraction by small holes. *Phys. Rev.* **66**, 163 (1944).
2. Mitrofanov, O., et al. Terahertz pulse propagation through small apertures. *Appl. Phys. Lett.* **79**, 907-909, 2001.
3. Khromova, I., et al. Splitting of magnetic dipole modes in anisotropic TiO2 microspheres. *Laser and Photonics Reviews* **10**, 4, 698 (2016).
4. Bhattacharya, A. et al. Large near-to-far field spectral shifts for terahertz resonances. *Phys. Rev. B* **93**, 035438 (2016).
5. A. Bitzer, A. et al. Terahertz near-field imaging of electric and magnetic resonances of a planar metamaterial. *Opt. Express* **17,** pp. 3826–3834 (2009).
6. Alonso-González, P. et al. Ultra-confined acoustic THz graphene plasmons revealed by photocurrent nanoscopy. *arXiv* 1601.05753 (2016).

7. Huber, A. J., Keilmann, F., Wittborn, J., Aizpurua, J. and Hillenbrand, R., Terahertz near-field nanoscopy of mobile carriers in single semiconductor nanodevices. *Nano Lett*. **8**, 3766–3770 (2008).

8. Jacob, R. et al. Intersublevel Spectroscopy on Single InAs-Quantum Dots by Terahertz Near-Field Microscopy. *Nano Lett*. **12**, 4336-4340 (2012).

9. Chen, H. et al. Performance of THz fiber-scanning near-field microscopy to diagnose breast tumors. *Optics Express* **19**, 19523-19531 (2011).

10. Ash, E. and Nicholls, G. Super-resolution aperture scanning microscope. *Nature* **237**, 510–512 (1972).

11. Mitrofanov, O. et al. Collection-mode near-field imaging with 0.5-THz pulses. *IEEE J. Sel. Top. Quantum Electron*. **7**, 600–607 (2001).

12. Moon, K. et al. Subsurface nanoimaging by broadband terahertz pulse near-field microscopy. *Nano Lett*. **15**, 549–552 (2015).

13. Dean, P. et al. Apertureless near-field terahertz imaging using the self-mixing effect in a quantum cascade laser. *Appl. Phys. Lett*. **108**, 091113 (2016).

14. Mitrofanov, O., Brener, I., Luk, T. S. and Reno, J. L. Photoconductive Terahertz Near-Field Detector with a Hybrid Nanoantenna Array Cavity. *ACS Photonics* **2**, 1763–1768 (2015).

15. Cocker, T. et al. A. An ultrafast terahertz scanning tunnelling microscope. *Nat. Photonics* **7**, 620–625 (2013).

16. Blanchard, F., Doi, A., Tanaka, T. and Tanaka, K. Real-Time, Subwavelength Terahertz Imaging, *Annu. Rev. Mater. Res*. **43**, 237–259 (2013).

17. Kawano, Y. and Ishibashi, K. An on-chip near-field terahertz probe and detector. *Nat. Photonics* **2**, 618-621 (2008).

18. Wade, C. G. et al. Real-Time Near-Field Terahertz Imaging with Atomic Optical Fluorescence. *Nature Photonics* **11**, 40–43 (2017).

19. Adam, A. Review of near-field terahertz measurement methods and their applications. *J. Infrared, Millimeter and Terahertz Waves*, **32,** 976-1019 (2011).

20. Huth, F. et al. "THz-TDS based near-field imaging and spectroscopy at 25 nm length scale," in *2015 European Conference on Lasers and Electro-Optics - European Quantum Electronics Conference*, (Optical Society of America, 2015), paper CC_5_2.

21. Bouwkamp, C. On Bethe's theory of diffraction by small holes. *J. Philips Res. Rep*. **5**, 321-332 (1950).

22. Viti, L. et al. Black Phosphorus Terahertz Photodetectors. *Adv. Mater.* **27**, 5567-5572 (2015).
23. Vitiello, M. S. et al. Room-temperature terahertz detectors based on semiconductor nanowire field-effect transistors. *Nano Lett.* **12**, 96–101 (2012).
24. Vitiello, M. S. et al. Semiconductor nanowires for highly sensitive, room-temperature detection of terahertz quantum cascade laser emission. *Appl. Phys. Lett.* **100**, 241101 (2012).
25. Vitiello, M. S. et al. One dimensional semiconductor nanostructures: An effective active-material for terahertz detection. *APL Mater.* **3**, 026104 (2015).
26. Vicarelli, L. et al. Graphene field-effect transistors as room-temperature terahertz detectors. *Nature Mater.* **11**, 865–871 (2012).
27. El Fatimy, A. et al. Epitaxial graphene quantum dots for high-performance terahertz bolometers. *Nature Nanotech.* **11**, 335-338 (2016).
28. Viti, L. et al. Efficient Terahertz detection in black-phosphorus nano-transistors with selective and controllable plasma-wave, bolometric and thermoelectric response. *Scientific Reports* **6**, 20474 (2016).
29. Viti, L. et al. Heterostructured hBN-BP-hBN Nanodetectors at Terahertz Frequencies. *Adv. Mater.* 28, 7390 -7396 (2016).
30. Dyakonov, M. and Shur, M. Detection, mixing, and frequency multiplication of terahertz radiation by two-dimensional electronic fluid. *IEEE Trans. Electron. Dev.* **43**, 380–387 (1996).
31. Ravaro, M. et al. Detection of a 2.8 THz quantum cascade laser with a semiconductor nanowire field-effect transistor coupled to a bow-tie antenna. *Appl. Phys. Lett.* **104**, 083116 (2014).
32. Boppel, S. et al. 0.25-µm GaN TeraFETs Optimized as THz Power Detectors and Intensity-Gradient Sensors. *THz Science and Technol., IEEE Trans.* 6, 348-350 (2016).
33. Boppel, S. et al. CMOS integrated antenna-coupled field-effect transistors for the detection of radiation from 0.2 to 4.3 THz. *IEEE Transactions on Microwave Theory and Techniques* **60**, 3834-3843 (2012).
34. Vitiello, M. S. and Tredicucci, A. Tunable Emission in THz Quantum Cascade Lasers. *IEEE Transactions on Terahertz Science and Technology* **1**, 76-84 (2011).

35. Castellano, F. et al. Tuning a microcavity-coupled terahertz laser. *Appl. Phys. Lett.* **107**, 261108 (2015).
36. Chen, L., Wei, Y., Zang, X., Zhu, Y., Zhuang, S. Excitation of dark multipolar plasmonic resonances at terahertz frequencies *Sci. Reports 6, 22027* (2016).
37. Mitrofanov, O. et al. Detection of internal fields in double-metal terahertz resonators, *Appl. Phys. Lett.*, in press (2017).

**Figure captions**

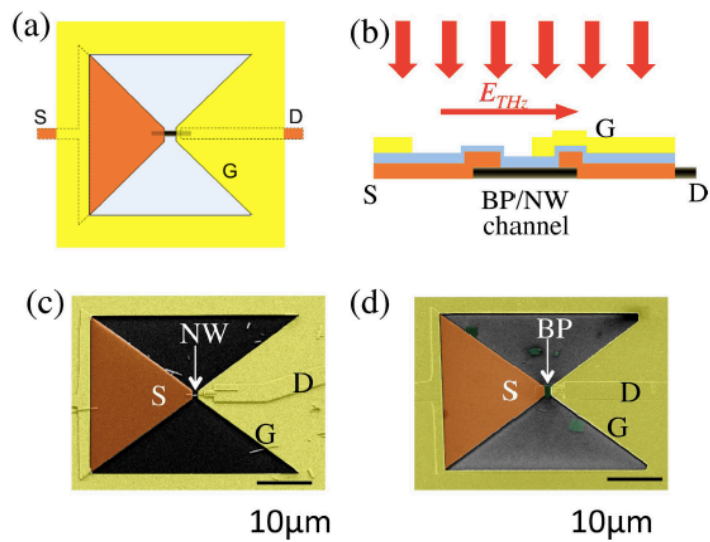

**Figure 1. Terahertz near-field probe with an embedded nano-detector**. Schematics of the near-field probe top view **(a)** and the cross section **(b)**. The electric field $E$ (red arrow in **b**) of the THz wave induces oscillating field between the source (S) and gate (G) electrodes. **c)** Scanning electron microscopy (SEM) image of a near-field probe with a 20 μm aperture and an embedded InAs NW as the FET channel. **d)** SEM image of a near-field probe with a 20 μm aperture and an embedded flake of black phosphorus (BP).

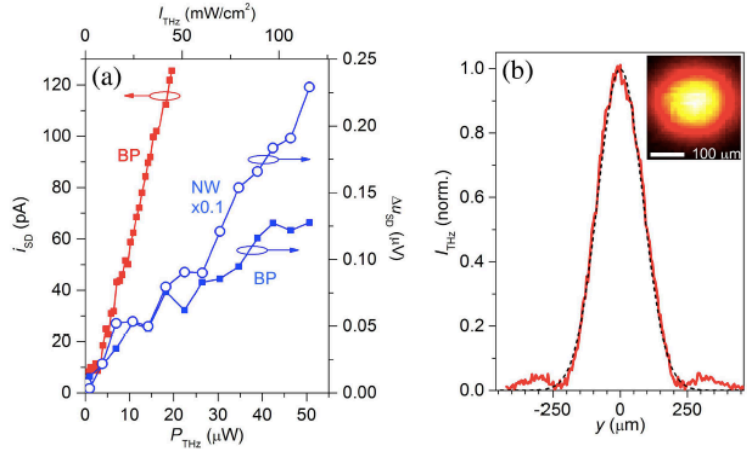

**Figure 2. Detection of 3.4 THz radiation by near-field probes with BP and NW detectors. a)** DC source-drain current $i_{SD}$ (photo-current mode) and source-drain voltage $\Delta u_{SD}$ (photovoltaic mode) in 20x20 μm² aperture probes as functions of the total power from the 3.4 THz QCL focused on the probe. **b)** The inset shows a THz image of the focused QCL beam, detected by the NW-detector in **a**. The main panel shows the THz beam intensity profile (red line) along the vertical axis measured by the NW-based detector with a 20 μm input aperture. The dashed black line shows a Gaussian fit $I_{THz} = I_0 \exp(-(y^2/(2r^2)))$, where $r = 83.5$ μm.

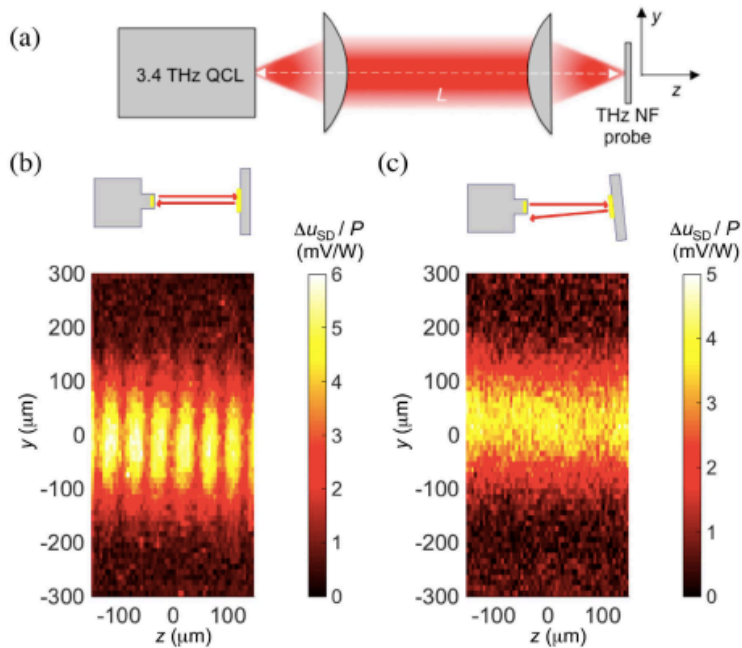

**Figure 3. Mapping of the 3.4 THz focused beam by the near-field probe with the BP detector. a)** Schematic of the experimental system: the QCL beam is focused the near-field probe using a pair Picarin lenses (focal lengths: 2 cm and 5 cm). The near-field probe is translated in the *yz*-plane. **b-c)** The THz intensity maps normalized to the incident THz power recorded by the 20 μm aperture BP probe with its surface oriented normally to the optical axis (**b**) and at a small angle (**c**).

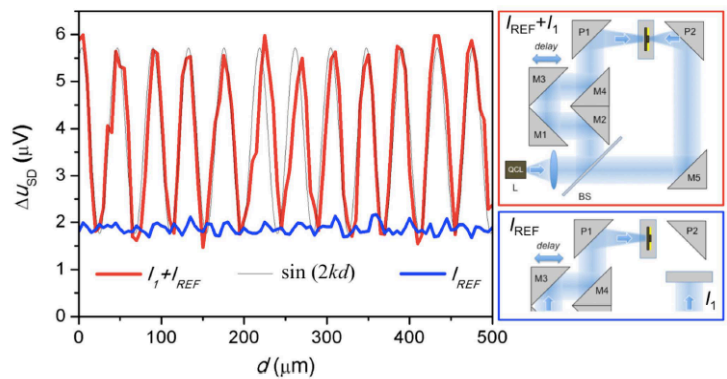

**Figure 4. Coherent detection using a near-field probe with an embedded NW detector.** Photovoltage as a function of the delay stage position (red). The photovoltage produced by the reference beam alone is shown as a blue trace for comparison. Insets show schematics of the experimental system for the two scans.

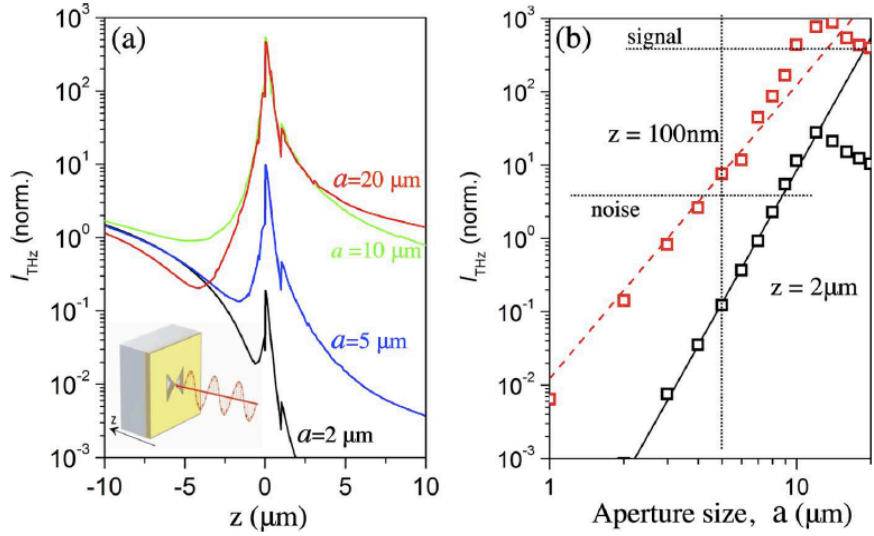

**Figure 5. Numerical simulation of the THz field in the aperture region. a)** THz intensity ($I_{THz} \propto |E|^2$, where $E$ is the electric field) along the optical axis passing through the center of 2, 5, 10 and 20 μm apertures and the source-gate gap of 500 nm. **b)** $I_{THz}$ at the detector position, $z = -100$ nm (red curve), for apertures ranging from 1-20 μm, compared to the $I_{THz}$ (black curve) at $z = -2$ μm. The red dashed line and the black solid line show the $a^4$ and $a^6$ power-law dependencies, respectively. The THz power in panels **a,b** is normalized to the incident field intensity. The dotted horizontal lines labeled signal and noise mark the experimental signal-to-noise ratio (SNR) corresponding to the simulated signal value at 20μm aperture (red square). The vertical dotted line set the minimum aperture size limit for SNR>1.

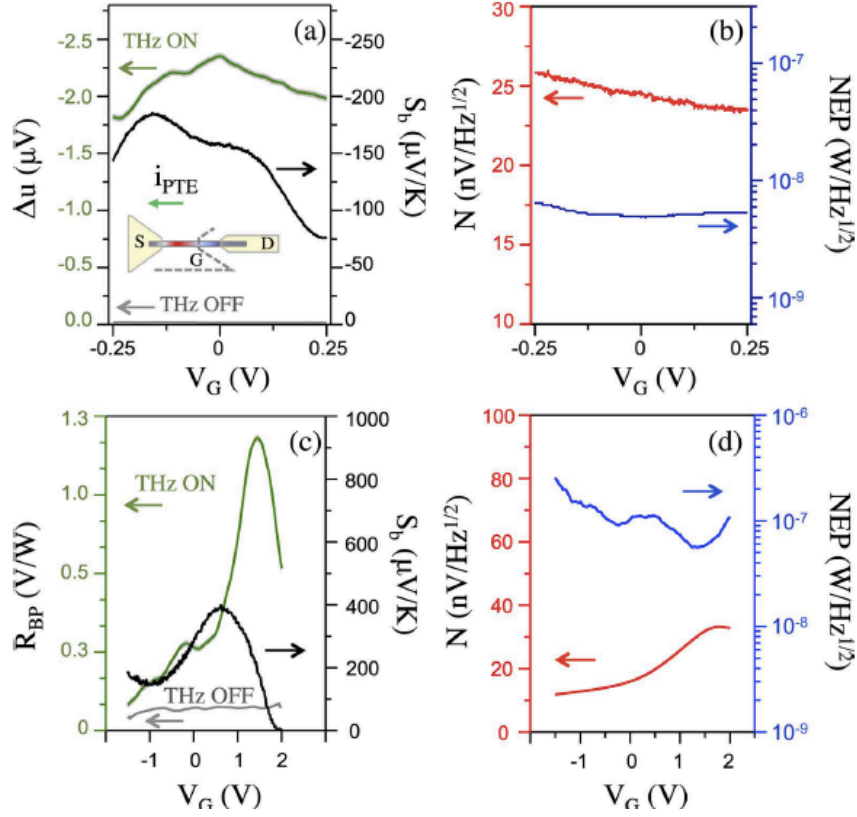

**Figure 6. Aperture-coupled THz FET nano-detector response as a function of gate voltage.** (a) left vertical axis: photovoltage $\Delta u_{SD}$ (THz beam on) and noise voltage (THz beam off) of sample A plotted as a function of the gate voltage $V_G$. Inset: schematics of the gated NW FET channel and thermoelectric current direction; right vertical axis: Seebeck coefficients determined from the theoretical Mott equation; (b) Johnson–Nyquist noise (left) and noise equivalent power (right) of sample A plotted as a function of the gate bias; (c) left vertical axis: Responsivity (THz beam on) and noise voltage (THz beam off) of sample B plotted as a function of the gate voltage $V_G$. right vertical axis: Seebeck coefficients; (d) Johnson–Nyquist noise (left) and noise equivalent power (right) of sample A plotted as a function of the gate bias.